\documentclass[prd,twocolumn,showpacs,a4paper]{revtex4}
\usepackage{graphicx}
\usepackage{amsfonts,amssymb}
\usepackage{amsmath}

\newcommand{\fud}{\frac{1}{2}}
\newcommand{\D}{{\mathcal{D}}}
\DeclareMathOperator{\Tr}{Tr}%
\DeclareMathOperator{\sign}{sign}%
\DeclareMathOperator{\PV}{P}
\DeclareMathOperator{\csch}{csch}

\newcommand{\expp}[1]{ \mathop\mathit{e}\nolimits^{#1}}

\newcommand{\derp}[3][]{\frac{\partial^{#1} #2}{\partial #3^{#1}}}

\newcommand{\derff}[3]{\frac{\delta^2 #1}{\delta #2 \, \delta #3}}
\newcommand{\av}[1]{\langle #1 \rangle}

\newcommand{\ud}[2][]{\textrm{d}^{#1}{#2}\,}
\newcommand{\uD}[1]{\D{#1}\,}
\newcommand{\vd}[2][]{\textrm{d}^{#1}{#2}}
\newcommand{\Eqref}[1]{Eq.~\eqref{#1}}
\newcommand{\ie}{\emph{i.e.}}

\renewcommand{\Re}{\mathop\mathrm{Re}\nolimits}
\renewcommand{\Im}{\mathop\mathrm{Im}\nolimits}
\newcommand{\vect}{\mathbf}

\newcommand{\TO}{^{\scriptscriptstyle (T=0)}}
\newcommand{\udpi}[2][]{\frac{\textrm{d}^{#1}{#2}}{(2\pi)^{#1}}}
\newcommand{\Lp}{L_\mathrm{Pl}}
\newcommand{\Mp}{M_\mathrm{Pl}}
\newcommand{\Sint}{S_\mathrm{int}}
\newcommand{\mth}{m_T}

\newcommand{\SigmaR}{\Sigma_\mathrm R}
\allowdisplaybreaks

\begin{document}

\title{Propagation in a thermal graviton background}

\author{Daniel Arteaga}\email{darteaga(at)ub.edu}
    \affiliation{Departament de F\'\i sica Fonamental, Universitat de Barcelona, Av.~Diagonal
    647, 08028 Barcelona, Spain}
\author{Renaud Parentani}\email{Renaud.Parentani(at)th.u-psud.fr}
    \affiliation{Laboratoire de Physique Th\'eorique, CNRS UMR 8627,
    Universit\'e Paris XI, 91405 Orsay Cedex, France}
\author{Enric Verdaguer}\email{verdague(at)ffn.ub.es}
    \affiliation{Departament de F\'\i sica Fonamental and CER en
    Astrof\'\i sica, F\'\i sica de Part\'\i cules i Cosmologia,
    Universitat de Barcelona, Av.~Diagonal 647, 08028 Barcelona,
    Spain}


\begin{abstract}
It is well known that radiative corrections evaluated in
nontrivial backgrounds lead to effective dispersion relations
which are not Lorentz invariant. Since gravitational interactions
increase with energy, gravity-induced radiative corrections could
be relevant for the trans-Planckian problem. As a first step to
explore this possibility, we compute the one-loop radiative
corrections to the self-energy of a scalar particle propagating in
a thermal bath of gravitons in Minkowski spacetime. We obtain
terms which originate from the thermal bath and which indeed break
the Lorentz invariance that possessed the propagator in the
vacuum. Rather unexpectedly, however, the terms which break
Lorentz invariance vanish in the high three-momentum limit. We
also found that the imaginary part, which gives the rate of
approach to thermal equilibrium, vanishes at one loop.
\end{abstract}

\maketitle

\section{Introduction}

In recent years modified dispersion relations which break Lorentz
invariance have appeared in different contexts of gravitational
physics.

On the one hand, they appeared in several works which address the
so-called \emph{trans-Planckian problem}
\cite{Jacobson91,Jacobson93,Jacobson99}. In black hole physics
 the modes responsible for Hawking radiation reach arbitrarily high
energies near the black hole horizon when measured by a
free-falling observer. This observation has lead some authors to
study the robustness of Hawking radiation when modifying (from the
outset) the dispersion relation beyond the Planck scale
\cite{Unruh95,BroutEtAl95,CorleyJacobson96,Helfer03}. A similar
problem appears in inflationary cosmology: the modes responsible
for the large scale structure had length scales much smaller than
the Planck length in the early stages of inflation. Therefore it
is of interest to determine to what extent the properties of the
fluctuation spectrum are sensitive to modifications of the
dispersion relation at the Planck scale
\cite{MartinBrandenberger00,Niemeyer00,NiemeyerParentani01,MartinBrandenberger03}.
In these works, the introduction of a nontrivial dispersion
relation has been originally suggested by formal analogies with
condensed matter
\cite{Unruh81,Unruh95,Visser98,GarayEtAl99,VisserEtAl01}. However,
the main physical motivation is that unknown effects of quantum
gravity (such as radiative corrections) might introduce nontrivial
dispersion relations in these contexts. Following 't Hooft's
\cite{tHooft96} observation that strong gravitational interactions
in the near horizon region might alter the semiclassical
description of black hole evaporation, a first dynamical
realisation of this line of sought has been pursued in
Refs.~\cite{Parentani01,Parentani02}.

On the other hand, there have been hints from high energy
astroparticle physics that we could already see the effects of a
violation of the Lorentz symmetry. In particular modified
dispersion relations have been used in order to provide an
explanation for observations of ultrahigh energetic cosmic rays
(with energies higher than $10^{20}\,\mathrm{eV}$) beyond the
Greisen-Zatsepin-Kuzmin (GZK) cutoff
\cite{TakedaEtAl98,Kifune99,Amelino-CameliaPiran01a,Amelino-CameliaPiran01b}.
As in the trans-Planckian problem, several dispersion relations
have been exploited, inspired by various approaches to quantum
gravity \cite{AlfaroEtAl99,AlfaroEtAl02}, string theory
\cite{KosteleckySamuel89}, non-conmutative field theory
\cite{CarrollEtAl01}, variation of couplings
\cite{KosteleckyEtAl02} and multiverses \cite{Bjorken03}. Modified
dispersion relations often break the energy degeneracy for a given
three-momentum: the particle energy can become helicity dependent
\cite{AlfaroEtAl99,AlfaroEtAl02} and light can become birefringent
\cite{CarrollEtAl90,KosteleckyMatthew01}. A popular form for these
modified dispersion relations below the Planck scale is, in the
cosmological rest frame,
\begin{equation} \label{Eq1}
    E^2 = m^2 + |\vect p|^2 + \sum_{n \ge 3} \eta_n
    \frac{|\vect p|^n}{\Mp^{n-2}},
\end{equation}
where $\Mp$ is Planck's mass and $\eta_n$ are coefficients of
order 1 if the Planck mass is the relevant scale. Jacobson
\emph{et al.}\ have studied the constraints on the possible values
of the parameters $\eta_n$ based on current astrophysical
observational data
\cite{JacobsonEtAl02,JacobsonEtAl03a,JacobsonEtAl03b,JacobsonEtAl03c}.
It is important to notice that both the real and imaginary parts
$\eta_3$ are already observationally constrained to be much
smaller than 1. This shows that the ``natural'' assumption that
the $\eta_n$ should be of order 1 is perhaps too naive. The
possibility of quadratic modifications of the dispersion relation
is also strongly constrained
\cite{CarrollEtAl90,KosteleckyMatthew01}.

As we learned from particle physics, Lorentz invariance is a key
element in renormalized quantum field theory
\cite{Hatfield,WeinbergQFT}. Nevertheless, it might be broken by
quantum gravity at a fundamental level, or, alternatively, it may
be broken in an effective way in nontrivial backgrounds only.
Indeed in backgrounds which possess a preferred reference frame,
radiative corrections to self-energies might contain terms which
effectively break the (local) Lorentz invariance ---\ie, the
Lorentz invariance in the tangent plane. Let us emphasize that
this second possibility does not imply any kind of fundamental
breaking of the Lorentz symmetry or any new physics. Rather by
``effective breaking of the Lorentz invariance'' we mean that
radiative corrections to the dispersion relation may contain terms
which depend on vector or tensor fields characterizing the
background.

In this paper we explore a model in the framework of this second
alternative. We consider the propagation of a scalar particle
immersed in a thermal bath of gravitons. Our aim is to determine
how the inertia of the thermal bath affects the propagation of the
scalar particle through radiative corrections. To this end, we
shall compute the thermal corrections to its self-energy. The real
part of the self-energy gives the thermal mass shift and might
introduce as well a nontrivial dispersion relation
\cite{DonoghueEtAl85}. The imaginary part instead gives the rate
of approach to thermal equilibrium \cite{Weldon83}.

In fact it is well known that quantum effects evaluated in
nontrivial backgrounds induce non Lorentz invariant terms in the
dispersion relation. For instance, the self-energy of a charged
fermion immersed in a thermal bath of photons has been computed in
Refs.~\cite{Weldon82,DonoghueHolstein83,DonoghueEtAl85,Weldon99},
and the effects of the thermal bath on the speed of light have
been studied in Refs.~\cite{Tarrach83,Barton90,LatorreEtAl95}. Let
us mention also that electromagnetic interaction in curved
spacetimes also leads to modifications of the dispersion
relations. The effect of QED vacuum polarization on the speed of
light in nontrivial spacetimes has been studied in
Refs.~\cite{DrummondHathrell80,Shore02,Shore02b,Shore03,Shore03b}.
However, most calculations done so far do not concern
gravitational interactions. Gravity-induced corrections should be
dominant at energies approaching the Planck scale, hence relevant
for the situations concerned with the trans-Planckian problem.
Additionally notice that gravitational interactions are universal
and not limited to charged particles.

Two important points concerning our approach should be mentioned.
On the one hand, since gravitational interactions are
non-renormalizable, the system of the scalar field coupled to
gravity should be conceived as an effective field theory
 \cite{Weinberg79,WeinbergQFT}. Although the full theory is
non-renormalizable, low-energy predictions which do not depend on
the Planck scale behavior of gravity can be extracted
 \cite{Donoghue94a,Donoghue94b,Donoghue95}. On the other hand,
since thermal corrections do not affect the ultraviolet properties
of the theory, thermal field theory fits well with the effective
theory approach. At a technical level, we shall incorporate the
thermal effects through the real-time description of thermal field
theory \cite{LandsmanWeert87,DasThermal,LeBellac}, which can be
seen as a particular application of the closed-time path (CTP)
method in field theory \cite{Schwinger61,Keldysh65,ChouEtAl85}.

We should also emphasize that we shall compute only the lowest
order corrections. Hence, we deal with free gravitons; \ie, we
neglect the backreaction of the scalar field on the metric
perturbations. The modifications to the metric propagator can be
computed both in the framework of stochastic gravity
\cite{CalzettaHu94,HuMatacz95,HuSinha95,CamposVerdaguer96,
CalzettaEtAl97,MartinVerdaguer99a,MartinVerdaguer99c,MartinVerdaguer00,HuVerdaguer03,
BarrabesFrolovParentani00} or in the large-$N$ limit of quantum
gravity \cite{Tomboulis77} (both approaches can be seen to be
equivalent \cite{RouraThesis,RouraVerdaguerInPrep}). A computation
along these lines (but in the semiclassical approximation) has
been carried out by Borgman and Ford \cite{BorgmanFord03}. They
study the fluctuations in the focusing of a bundle of geodesics,
propagating in a spacetime with metric perturbations induced by a
thermal scalar field. To our knowledge, the higher order
corrections to the self-energy have not yet been computed in the
quantum field theoretical framework.

Gravity-mediated modifications of the dispersion relation have
been also considered by Burgess \emph{et
al.}~\cite{BurgessEtAl03}. Working in the context of brane-world
scenarios, these authors compute the change to the dispersion
relation of photons and fermions generated by interaction with
four-dimensional effective gravitons. In that case the primary
source of violation of the Lorentz symmetry is the modification of
the dispersion relation of effective four-dimensional gravitons
induced by certain extra-dimensional configurations. Similarly to
our case, this modification is communicated to brane-bound
particles through gravitational interactions.

The plan  of the paper is the following. In Sec.~II we introduce
the action for the system considered ---namely, a scalar field
coupled to gravity. In Sec.~III we compute perturbatively the
leading correction to the self-energy of the scalar field at zero
temperature. In Sec.~IV we assume that the scalar field and the
gravitons are in thermal equilibrium at a given temperature, and
compute the corrections to the retarded self-energy. Finally in
Sec.~V we analyze the results and discuss the physical
significance of the terms which break Lorenz invariance. The
Appendixes contain reference material as well as some technical
details of the calculation. In Appendix \ref{app:FeyRules} we
present the Feynman rules for the scalar field coupled to
linearized gravity. In Appendix \ref{app:Int} we give the results
of the calculation of some integrals in dimensional
regularization. In Appendix \ref{app:TFT} we give a brief account
of the real-time approach to thermal field theory. Appendix
\ref{app:ABCD} is mostly technical and is devoted to the
calculation of certain integrals that appear in the calculation of
the thermal contribution to the self-energy.

Throughout this paper we shall use a system of units with
$\hbar=c=k_\mathrm B=1$. The signature of the metric will be
$(-,+,+,+)$.

\section{System}

We consider a minimally coupled real scalar field $\phi$ of mass
$m$ propagating in a spacetime characterized by a metric
$g_{\mu\nu}$. The action for the field is
\begin{subequations}
\begin{equation}
    S_{\phi,g} = - \int \ud[4]x \sqrt{-g} \left( \fud g^{\mu\nu}
    \partial_\mu \phi\, \partial_\nu \phi + \fud m^2 \phi^2
    \right),
\end{equation}
and the action for the metric is
\begin{equation}
    S_g =  \frac{2}{\kappa^2} \int \ud[4]x \sqrt{-g}\, R,
\end{equation}
\end{subequations}
where $R$ is the Ricci scalar, $g$ is the determinant of the
metric, and $\kappa = \sqrt{32 \pi G} = \sqrt{32\pi}\, \Lp$ is the
gravitational coupling constant, with $G$ being Newton's constant
and $\Lp$ being Planck's length. Assuming that the metric is a
small perturbation of Minkowski spacetime, $g_{\mu\nu} =
\eta_{\mu\nu} + \kappa h_{\mu\nu}$, the complete action $S =
S_{\phi,g} + S_g$ can be decomposed into the free scalar field,
graviton  and interaction actions as $S = S_\phi + S_h + \Sint$.
These actions, expanded in powers of $\kappa$, are
\begin{subequations}
\begin{align}
    S_\phi &=  \int \ud[4]x  \left( - \fud
    \partial_\mu \phi \, \partial^\mu \phi - \fud m^2 \phi^2
    \right), \label{ActPhiClas}
    \\
    \begin{split}
    S_h &= \int \ud[4]x \bigg(- \frac12 \partial^\alpha h^{\mu\nu} \partial_\alpha h_{\mu\nu}
    +  \partial_\nu h^{\mu\nu} \partial^{\alpha}
    h_{\mu\alpha}
    \\
    &\ \phantom{ = \int \ud[4]x \Big( } - \partial_\mu h \, \partial_\nu h^{\mu\nu}
    + \frac12 \partial^\mu h\, \partial_\mu h \bigg) + O(\kappa),
    \end{split}\label{GravProp}\\
    \Sint &=  \int \ud[4]x \left( \frac\kappa2 T^{\mu\nu} h_{\mu\nu} +
    \frac{\kappa^2}{4} U^{\mu\nu\alpha\beta} h_{\mu\nu}
    h_{\alpha\beta} \right)
     +  O(\kappa^3),
    \label{IntTerm}
\end{align}
\end{subequations}
where $h=h^{\mu}_{\phantom\mu\mu}$, $T_{\mu\nu}$ is the stress
tensor of the scalar field,
\begin{equation}
    T_{\mu\nu}  =  \partial_\mu \phi \partial_\nu \phi - \fud
    \eta_{\mu\nu} \partial_\alpha \phi \partial^\alpha \phi- \fud \eta_{\mu\nu} m^2
    \phi^2,
\end{equation}
and
\begin{equation}
\begin{split}
    U_{\mu\nu\alpha\beta} = &- 2\eta_{\nu\alpha} \partial_\mu \phi
    \partial_\beta \phi + \eta_{\mu\nu} \partial_\alpha \phi
    \partial_\beta \phi \\
    &+\left(\frac12 \eta_{\mu\alpha} \eta_{\nu\beta} - \frac14
    \eta_{\mu\nu} \eta_{\alpha\beta}\right)(\partial^\sigma\phi
    \partial_\sigma\phi + m^2\phi^2).
\end{split}
\end{equation}
Indices are raised and lowered with the background metric
$\eta_{\mu\nu}$. We have kept only the free terms in the action
for the gravitons $S_h$ because we shall only compute the lowest
order corrections of the self-energy of the $\phi$ field.

To compute these corrections we need to introduce counterterms in
order to cancel divergences. Since our system is
non-renormalizable, it has to be understood as an effective field
theory, a low-energy approximation of a more fundamental theory at
the Planck scale
\cite{Donoghue94a,Donoghue94b,Donoghue95,WeinbergQFT}. In order to
compute to a given precision $E^n \kappa^n$, where $E$ is the
energy of the process, one has to introduce all possible
counterterms compatible with the symmetry whose coefficients are
of order $\kappa^n$ at most. In our case the most general action
for the counterterms for the scalar field action up to order
$\kappa^2$ which is compatible with the Poincar\'e symmetry is
\begin{equation}
\begin{split}
    S_\mathrm{count} =  - \int \ud[4]x  \bigg[ &\ \fud (m_0^2-m^2)
    \phi^2 \\
     + &\ \fud (Z -1) (\partial_\mu\phi \partial^\mu\phi + m^2
     \phi^2)
    \\
     +
    &\ \frac14 \kappa^2 C_0
    (\partial_\mu \partial^\mu \phi)^2 \bigg] + O(\kappa^4),
\end{split}
\end{equation}
where $m_0=m+O(\kappa^2)$ is the bare mass, $Z =1+O(\kappa^2)$ is
the field renormalization parameter and $C_0=C+O(1)$ is a bare
four-derivative coefficient. The finite coefficient $C$ is \emph{a
priori} unknown and constitutes an external input of the theory.
The value of $C$ should be determined by experiments or by
knowledge of the underlying more fundamental theory. On the other
hand, we do not introduce counterterms to the interaction action
because we shall not compute vertex corrections.

Additionally the graviton action \eqref{GravProp} must be
supplemented with a gauge-fixing term. We will work with the
harmonic gauge $\partial_\nu h^{\mu\nu} - (1/2)\partial^\mu h =
0$, whose appropriate gauge-fixing action is
\cite{tHooftVeltman74,Veltman76}
\begin{equation}
    S_\mathrm{gf} = - \int \ud[4] x \left(\partial_\nu h^{\mu\nu} -
    \fud \partial^\mu h \right) \left(\partial^\lambda h_{\mu\lambda} -
    \fud \partial_\mu h \right).
\end{equation}
No Faddeev-Popov ghost fields are needed since we will not
consider graviton self-interactions.

\section{Zero temperature}

The aim in this section is to compute the leading contribution to
the vacuum self-energy of the scalar particle. At zero temperature
the self-energy $\Sigma\TO(p^2)$ is related to the Feynman
propagator $G\TO_\mathrm{F}(p)$ through
\begin{equation}\label{Feynman}
    G\TO_\mathrm{F}(p) = \frac{-i}{p^2+m^2+\Sigma\TO(p^2)}.
\end{equation}
We recall that the self-energy can be computed as the sum of all
one-particle irreducible diagrams with amputated external legs
\cite{Peskin,WeinbergQFT,Hatfield}. In order to regulate
divergences appearing in the calculation we will use dimensional
regularization \cite{Leibbrandt75,TarrachPascual}.

\begin{figure}
    \centering\hfill
    \includegraphics{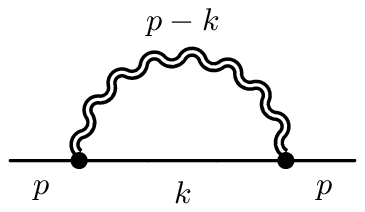}\hfill
    \includegraphics{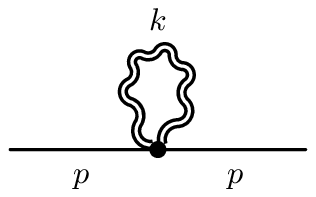}\hfill\mbox{}
    \caption{The two Feynman diagrams needed
    for the calculation of the self-energy: respectively, $\Sigma_{(1)}(p)$ and $\Sigma_{(2)}(p)$.}
    \label{fig:FeyDiag}
\end{figure}

The two diagrams which may contribute to order $\kappa^2$ are
shown in Fig.~\ref{fig:FeyDiag}. One must also take into account
the contribution of the counterterms:
\begin{equation}\label{Sigma0}
\begin{split}
    \Sigma\TO(p^2) = &\ (m_0^2-m^2) + (Z-1) (p^2+m^2) + \kappa^2 C_0 p^4 \\ &\ + \Sigma\TO_{(1)}(p^2) + \Sigma\TO_{(2)}(p^2) +
    O(\kappa^4).
\end{split}
\end{equation}
We first concentrate on the first diagram $\Sigma\TO_{(1)}(p^2)$.
Applying the Feynman rules described in Appendix
\ref{app:FeyRules} we find, in $d$ spacetime dimensions,
\begin{equation}
\begin{split}
    -i\Sigma\TO_{(1)}(p^2)
    =  \mu^\varepsilon\int  &\ \frac{\vd[d]k}{(2\pi)^d}  \tau_{\mu\nu} (p,k) \tau_{\alpha\beta} (k,p)\\
     \times  &\ \frac{-i \mathcal P^{\mu\nu\alpha\beta}}{(p-k)^2-i\epsilon}
    \frac{-i}{k^2+m^2-i\epsilon} ,
\end{split}
\end{equation}
where $\varepsilon=4-d$, $\mathcal P_{\mu\nu\alpha\beta}$ and
$\tau_{\mu\nu}(p,k)$ are given in Appendix \ref{app:FeyRules}, and
$\mu$ is an arbitrary mass scale. Developing the products in the
numerator we find
\begin{equation} \label{SigmaDecInt}
\begin{split}
    -i\Sigma\TO_{(1)}(p^2) = &\
     \frac{\kappa^2}{2}
    p^2 \eta^{\mu\nu} I_{\mu\nu}(p) - \kappa^2 m^2 p^\mu
    I_{\mu}(p) \\ &- \kappa^2 m^4 (1+\varepsilon/4)I(p) +
    O(\varepsilon),
\end{split}
\end{equation}
where the momentum integrals $I(p)$, $I_\mu(p)$ and
$I_{\mu\nu}(p)$ are defined and  computed in
Appendix~\ref{app:Int}. The result for $\Sigma\TO_{(1)}(p^2)$ is
\begin{equation}
\begin{split}
    \Sigma\TO_{(1)}(p^2) = &\ \frac{\kappa^2}{(4\pi)^2}  \bigg[
    \frac{m^4}{\hat\varepsilon} +  {2m^4}
    +\frac{m^2p^2}{\hat\varepsilon} + 2m^2p^2 \\
    & -\left( \frac{ m^6}{2p^2} +\frac{m^4}{2}\right) \ln\left( 1 + \frac{p^2}{m^2}-i\epsilon\right) \\
    & - \left( m^4 + m^2p^2 \right)
    \ln\left(\frac{p^2+m^2}{\mu}-i\epsilon\right) \bigg] + O(\varepsilon),
\end{split}
\end{equation}
where ${1}/{\hat\varepsilon} = {2}/{\varepsilon} - \gamma +
    \ln 4\pi$.
The second diagram $\Sigma\TO_{(2)}(p^2)$ is a massless tadpole,
and these are identically zero in dimensional regularization
\cite{Leibbrandt75}.

Once we include the counterterms and take the limit $\varepsilon
\to 0$ the renormalized self-energy is found to be
\begin{equation}\label{SigmaM}
\begin{split}
    \Sigma\TO(p^2) &=
    - \frac{\kappa^2}{(4\pi)^2} \left( \frac{ m^6}{2p^2} + \frac{m^4}{2} \right) \ln\left( 1 +
    \frac{p^2}{m^2}-i\epsilon \right) \\
     &\quad- \frac{\kappa^2}{(4\pi)^2} \left(m^4 + m^2p^2 \right)
    \ln\left(\frac{p^2+m^2}{\mu^2}-i\epsilon\right) \\
    &\quad + C \kappa^2 (p^2+m^2)^2 +
    O(\kappa^4),
\end{split}
\end{equation}
where we have absorbed divergences in the following way:
\begin{subequations}
\begin{align}
    m_0^2 &= m^2 - C \kappa^2 m^4 + O(\kappa^4)\\
    Z &= 1 + 2 C \kappa^2m^2- \frac{\kappa^2m^2}{(4\pi)^2} \left( \frac{1}{\hat\varepsilon} + 2     \right)+ O(\kappa^4),\\
    C_0 &= C + O(\kappa^4).
\end{align}
\end{subequations}
In order to relate the bare and the renormalized parameters we
have imposed the on shell renormalization condition
\begin{equation} \label{RenorCond}
    \Sigma\TO(-m^2) = 0,
\end{equation}
which fixes the position of the pole to be $m^2$. The complete
specification of the on shell renormalization scheme also involves
fixing the residue of the pole, which implies an additional
condition on the derivative of the self-energy
\cite{WeinbergQFT,Hatfield}. However this second condition is
found to be singular due to the divergence of the logarithms when
evaluated on shell. This kind of singularity is generic for
theories with interacting massless particles because of the
absence of well defined asymptotic regions in the presence of
long-range forces. For a discussion of this point in the case of
QED see Ref.~\cite{Hatfield}. Anyway in the paper we are just
interested position of the pole, which is always well defined.

Note that the only divergence arises from the field
renormalization. In particular, the four-derivative coefficient
$C$ does not get renormalized: there are no logarithmic
corrections to the $p^4$ term in the self-energy \eqref{SigmaM},
and $C$ coincides with its classical value $C_0$. Therefore it
would have been consistent not to include the four-derivative
counterterm and simply take $C=0$ from the beginning.

The self-energy develops a negative imaginary part for $-p^2>m^2$,
\begin{equation} \label{ImSigma}
\begin{split}
    \Im \Sigma\TO (p^2)=\theta(-p^2-m^2)\frac{\kappa^2}{16\pi} \left( \frac{
    m^6}{2p^2}+
    \frac{3m^4}{2} + m^2p^2 \right) \!,
\end{split}
\end{equation}
which, according to the optical theorem \cite{Peskin}, accounts
for the probability of a scalar particle with momentum $p$ to
decay into an on shell scalar and an on shell graviton. Notice
that on shell ($p^2=-m^2$) the imaginary part of the self-energy
vanishes, because there is no phase space available for the
spontaneous emission of a graviton.

\section{Finite temperature}

In this section we compute the thermal contribution to the
self-energy working within the real-time approach to thermal field
theory. In this approach, which is briefly explained in Appendix
\ref{app:TFT}, the number of degrees of freedom is doubled, and
one has to consider four propagators organized in a $2\times2$
matrix $G_{ab}(p)$.  The self-energy also becomes a matrix
$\Sigma^{ab}(p)$.  We will concentrate on the retarded propagator
$G_\mathrm R(p)=G_{11}(p)-G_{12}(p)$ for the following reasons:
First, it is the one that exhibits simple analytical properties at
finite temperature (analyticity in the upper $p^0$ plane);
\pagebreak second, as shown in Appendix  C, it is directly
connected with the retarded self-energy $\Sigma_\mathrm
R=\Sigma^{11}(p)+\Sigma^{12}(p)$ through
\begin{equation}
    G_\mathrm R(p) = \frac{-i}{p^2+m^2+\Sigma_\mathrm R(p)};
\end{equation}
and, finally, the position of its poles have well defined
interpretations in terms of energies and thermalization rates.
Furthermore, the retarded propagator is the one that is naturally
obtained from analytic continuation from the Euclidean propagator
in the imaginary-time formalism. For more details we refer to
Appendix \ref{app:TFT} and to
Refs.~\cite{LandsmanWeert87,DasThermal,LeBellac}. A quantum
mechanical model where this point will be studied in detail will
be presented elsewhere \cite{ArteagaParentaniVerdaguerInPrep}.

\subsection{Self-energy: Real part}

Let us proceed to the calculation of the real part of
$\Sigma_\mathrm R(p)$. Since $\Re \Sigma_\mathrm R(p) = \Re
\Sigma^{11}(p)$ [see Eq.~\eqref{SigmaR11}], we shall compute the
real part of $\Sigma^{11}(p)$ instead. We have to consider both
diagrams sketched in Fig.~\ref{fig:FeyDiag}, where now internal
propagators are taken to be thermal and of type 11:
\begin{equation}\label{Sigma11}
\begin{split}
    \Sigma^{11}(p) = &\ (m_0^2-m^2) + (Z - 1) (p^2+m^2) + \kappa^2 C_0 p^4 \\ &\ + \Sigma^{11}_{(1)}(p) + \Sigma^{11}_{(2)}(p) +
    O(\kappa^4).
\end{split}
\end{equation}
The first diagram, $\Sigma^{11}_{(1)}(p)$, is given by
\begin{widetext}
\begin{equation}
\begin{split}
    \Sigma_{(1)}^{11}(p) = \frac{i\kappa^2}{2} \mu^\varepsilon\int \udpi[d]{k}
     & \ \left[ \frac{1}{(p-k)^2-i\epsilon} + 2\pi
    i n(|p^0-k^0|) \delta\boldsymbol((p-k)^2\boldsymbol) \right] \\
    \times & \ \left[ \frac{1}{k^2+m^2-i\epsilon} + 2\pi
    i n(|k^0|) \delta(k^2-m^2) \right] g_1(p^2,k^2,p\cdot
    k)
\end{split}
\end{equation}
\end{widetext}
where
\begin{equation}
\begin{split}
    g_1(p^2,k^2,p \cdot k) &= -\frac{2}{\kappa^2}
    \tau_{\mu\nu}(p,k) \tau_{\alpha\beta}(k,p)
    \mathcal P^{\mu\nu\alpha\beta}\\
    &= k^2 p^2 - 2 (k \cdot p) m^2 -2(1+\varepsilon/4) m^4,
\end{split}
\end{equation}
the function $n(E)$ is the Bose-Einstein distribution,
\begin{equation}
    n(E) = \frac{1}{1-\expp{E/T}},
\end{equation}
and we recall that $\varepsilon=4-d$ and that $\tau_{\mu\nu}(p,k)$
and $\mathcal P^{\mu\nu\alpha\beta}$ are given in
Appendix~\ref{app:FeyRules}. On the other hand, at finite
temperature the tadpole diagram $\Sigma^{11}_{(2)}(p)$ no longer
vanishes since its temperature-dependent part  gives a finite
contribution,
\begin{equation}
    \Sigma_{(2)}^{11}(p) =- \frac{\kappa^2}{2} g_3(p^2) \int \udpi[4]k 2\pi \delta(k^2)
    n(|k^0|),
\end{equation}
where
\begin{equation}
\begin{split}
    g_3(p^2) = -\frac{2i}{\kappa^2} V_{\mu\nu\alpha\beta}(p,p) \mathcal
    P^{\mu\nu\alpha\beta} = 10m^2 + 4p^2,
\end{split}
\end{equation}
with $V_{\mu\nu\alpha\beta}(p,k)$ given in
Appendix~\ref{app:FeyRules}.

To order $\kappa^2$ one has
\begin{equation} \label{ReSigmaABC}
\begin{split}
    \Re \Sigma^{11}(p)
    &= \Re \Sigma^{\scriptscriptstyle (T=0)}(p^2)   -
    \frac{\kappa^2}{2} \left[ A(p) + B(p) + C(p^2) \right]\\ &\quad+ O(\kappa^4),
\end{split}
\end{equation}
where $\Sigma^{\scriptscriptstyle (T=0)}(p^2)$ is the $T=0$
self-energy [see Eqs.~\eqref{Sigma0} and \eqref{SigmaM}] and where
the integrals $A(p)$, $B(p)$ and $C(p^2)$ are defined through
\begin{subequations}
\begin{align}
\begin{split}
    A(p) &= \int \frac{\mathrm d^4 k}{(2\pi)^3} n(|p^0 - k^0|)
    \delta\boldsymbol( (p-k)^2 \boldsymbol) \\ &\qquad\times g_1(p^2,k^2,p \cdot k) \PV \frac{1}{k^2+m^2} \label{A},
\end{split}\\
\begin{split}
    B(p) &= \int \frac{\mathrm d^4 k}{(2\pi)^3}  n(|k^0|)
    \delta(k^2+m^2) \\&\qquad\times g_1(p^2,k^2,p \cdot k) \PV \frac{1}{(p-k)^2} ,  \label{B}
\end{split}\\
C(p^2) &= g_3(p^2) \int \frac{\mathrm d^4 k}{(2\pi)^3} n(|k^0|)
\delta(k^2), \label{C}
\end{align}
\end{subequations}
with ``$\PV$'' meaning principal value. We have set $d=4$ because
thermal contributions are ultraviolet finite. Note that the same
renormalization process which makes the $T=0$ self-energy finite
also renormalizes the $T>0$ self-energy; there is no need to
introduce additional temperature-dependent counterterms. Hereafter
we will concentrate in the on shell results, $p^0 = E_\vect p =
\sqrt{m^2+|\vect p|^2}$, since these are the ones we will need
afterwards for the calculation of the thermal mass and the
modified dispersion relation.

The integrals $A(p)$ and $C(p^2)$ take into account the effect of
the thermal gravitons on the scalar particle. In the on shell case
it is possible to give an explicit expression for these integrals
valid at any temperature (the details of the calculation can be
found in Appendix~\ref{app:ABCD}):
\begin{align}
    A(E_\vect p, \vect p)&= -\frac16 m^2 T^2, \label{ResultA}\\
    C(-m^2) &= \frac12 m^2 T^2. \label{ResultC}
\end{align}
Notice that $A(E_\vect p, \vect p)$ does not depend on the
three-momentum $\vect p$. This is a surprise since Eq.~\eqref{A}
is not manifestly Lorentz invariant. This simplification, which is
also found in electrodynamics \cite{DonoghueEtAl85}, will have
important consequences  when computing the modified dispersion
relation.

The integral $B(p)$ takes into account the effect of the thermal
scalars in the heat bath. For temperatures well below the mass $m$
there are almost no scalar particles in the bath, and this indeed
shows up in an exponential suppression of $B(E_\vect p, \vect p)$
at low temperatures, $T \ll m$:
\begin{equation} \label{ResultBLowT}
\begin{split}
    B(E_\vect p, \vect p) \approx \sqrt{\frac{ m^5 T^3}{2\pi^3}} \left(
    \frac{m^2+2|\vect p|^2}{3m^2+4|\vect p|^2}\right) \expp{-m/T},
    \quad T \ll m.
\end{split}
\end{equation}
For temperatures of the order of the mass $m$, there are thermal
scalar particles in the bath and they give a significant
contribution to the self-energy. At high temperatures  $T \gg m$,
the leading contribution to the integral is given by
\begin{widetext}
\begin{equation} \label{ResultBHighT}
\begin{split}
    B(E_\vect p, \vect p) &\approx
    \frac{m^2 T^2\sqrt{m^2+ |\vect p|^2}}
{24|\vect p|}
 \ln \left(\frac{ 2\sqrt{m^2+ |\vect
    p|^2}- |\vect p|}{ 2\sqrt{m^2+ |\vect
    p|^2}+ |\vect p|}\right)  + \frac{1}{8} m^2 T^2, \quad T \gg m.
\end{split}
\end{equation}
The details of the calculation for both the low- and
high-temperature limits are given in Appendix~\ref{app:ABCD}.

Summarizing, according to Eqs.~\eqref{ReSigmaABC} and
\eqref{ResultA}--\eqref{ResultBHighT} the real part of the on
shell self-energy in the low- and high-temperature regimes is
given by:
\begin{equation} \label{ReSigmaT}
    \Re \Sigma_\mathrm R (E_\vect p, \vect p)
    \approx
    \begin{cases}
         -\dfrac{1}{6} \kappa^2 m^2 T^2
        -   \sqrt{\dfrac{m^5 T^3}{{8\pi^3}}}\kappa^2  \expp{-m/T} \left(
        \dfrac{m^2+2|\vect p|^2}{3m^2+4|\vect p|^2}\right) ,
        & T \ll m,\\[3ex]
         \dfrac{1}{48} \kappa^2 m^2 T^2 \left[  - 11
        + \dfrac{ \sqrt{m^2+ |\vect p|^2}}
     {|\vect p|}
         \ln \left(\dfrac{ 2\sqrt{m^2+ |\vect
       p|^2}+ |\vect p|}{ 2\sqrt{m^2+ |\vect
        p|^2}- |\vect p|}\right)\right] , & T \gg m.
    \end{cases}
\end{equation}
Notice that from the high-temperature result we can deduce that
for massless particles the on shell self-energy is exactly zero.
\end{widetext}

\subsection{Self-energy: Imaginary part}

Now we want to compute $\Im \Sigma_\mathrm R(p)$. Similarly to the
previous subsection we could compute $\Sigma^{11}(p)$ and then
make use of Eq.~\eqref{SigmaR11}; however, it is somewhat easier
for us to compute $\Sigma^{12}(p)$ and $\Sigma^{21}(p)$ and make
use of relation \eqref{cutR} instead. The self-energy
$\Sigma^{12}(p)$ is given by
\begin{equation}
\begin{split}
    \Sigma^{12}(p) &= i \frac{\kappa^2}{2} \int \frac{\mathrm d^4 k}{(2\pi)^2} n(k^0) \sign(k^0) \\
    &\quad \times n(p^0-k^0) \sign(p^0-k^0) \\
    &\quad \times \delta\boldsymbol( (p-k)^2 \boldsymbol) \delta(k^2+m^2) g_1(p^2,k^2,p \cdot
    k),
\end{split}
\end{equation}
where we used the property $\theta(-p^0) + n(|p^0|) =
\sign(p^0)n(p^0)$. In a similar way we find for $\Sigma^{21}$:
\begin{equation}
\begin{split}
    \Sigma^{21}(p) &= i \frac{\kappa^2}{2} \int \frac{\mathrm d^4 k}{(2\pi)^2}[1+ n(k^0)] \sign(k^0) \\
    &\quad \times [1+n(p^0-k^0)] \sign(p^0-k^0) \\
    &\quad \times \delta\boldsymbol( (p-k)^2 \boldsymbol) \delta(k^2+m^2) g_1(p^2,k^2,p \cdot
    k),
\end{split}
\end{equation}
where now we used $\theta(p^0) + n(|p^0|) = \sign(p^0)[1+n(p^0)]$.
Thus from Eq.~\eqref{cutR} and the two previous equations we get
\begin{equation} \label{ImSigmaR}
    \Im \Sigma_\mathrm R(p) = -\frac{\kappa^2}{4} D(p),
\end{equation}
where $D(p)$ is defined by
\begin{equation} \label{D}
\begin{split}
    D(p) &=  \int \frac{\mathrm d^4 k}{(2\pi)^2}F(p^0,k^0) g_1(p^2,k^2,p \cdot
    k)
    \\ &\quad \times\delta\boldsymbol( (p-k)^2 \boldsymbol) \delta(k^2+m^2)
\end{split}
\end{equation}
with
\begin{equation}
    \begin{split}
    F(p^0,k^0) &=  \sign(k^0)\sign(p^0-k^0) \\ &\quad \times \left[ 1 + n(k^0)
    + n(p^0-k^0)  \right],
    \end{split}
\end{equation}
which can be developed to give
\begin{equation}
    F(p^0,k^0)= \fud
    \sinh\left(\frac{p^0}{2T}\right)\csch\left(\frac{|p^0-k^0|}{2T}\right)\csch\left(\frac{|k^0|}{2T}\right).
\end{equation}
After manipulating the integral $D(p)$ (the details can be found
in Appendix~\ref{app:ABCD}) the imaginary part of the self-energy
can be expressed as the following phase-space integral:
\begin{equation} \label{IntImSigma}
    \Im \Sigma_\mathrm R(p) = \frac{\kappa^2m^2(m^2+2p^2)}{32\pi |\vect{p}|} \left| \int_{Q_1}^{Q_2}  \ud Q
    F(p^0,Q)\right|.
\end{equation}

Let us now evaluate the integral in \Eqref{IntImSigma} for an
arbitrary temperature. For simplicity, we restrict ourselves to
the case $p^0 > |\vect p|$, but including both $(p^0)^2 > m^2 +
|\vect p|^2$ and $(p^0)^2 < m^2 + |\vect p|^2$. In this situation
the integral can be performed analytically to give
\begin{widetext}
\begin{equation} \label{ImSigmaT}
    \Im \Sigma_\mathrm R(p) = \frac{\kappa^2T m^2(m^2+2p^2)}{32\pi |\vect p|}
    \ln \left[\frac{
    \sinh \left(\frac{(p^0+|\vect p|)^2+m^2}{4T(p^0+|\vect p|)}\right)
    \sinh \left(\frac{(p^0)^2-|\vect p|^2-m^2}{4T(p^0-|\vect p|)}\right)}
    {\sinh \left(\frac{(p^0-|\vect p|)^2+m^2}{4T(p^0-|\vect p|)}\right)
    \sinh \left(\frac{(p^0)^2-|\vect p|^2-m^2}{4T(p^0+|\vect p|)}\right)}
    \right].
\end{equation}
\end{widetext}
Making use of the property
\[
    \ln (\sinh x) \approx |x| - \ln 2 - i \pi \theta(-x) - \expp{-2|x|} ,  \quad
    |x| \gg 1,
\]
we easily obtain the low-temperature approximation
\begin{equation}
\begin{split}
     \Im \Sigma_\mathrm R(p) = &\ \theta(-p^2-m^2) \frac{\kappa^2}{16\pi}
      \left( \frac{m^6}{2p^2} + \frac{3m^4}{2} + m^2 p^2
    \right)\\
    &+O(\expp{-m/T}),
\end{split}
\end{equation}
which is in agreement with the zero-temperature result of
Eq.~\eqref{ImSigma}. At zero temperature the imaginary part of the
self-energy gives a net decay rate, but at finite temperature it
is related to the rate $\Gamma$ at which a particle approaches
thermal equilibrium \cite{Weldon83}:
\begin{equation} \label{Gamma}
    \Gamma = \Gamma_\mathrm d - \Gamma_\mathrm c = -\frac{1}{p^0} \Im \Sigma_{\mathrm R}.
\end{equation}
In this expression $\Gamma_\mathrm d$ and $\Gamma_\mathrm c$ are,
respectively, the annihilation and creation rates of particles
with energy $p^0$, which take into account stimulated absorbtion
and emission by the thermal bath. An arbitrary ensemble of
particles with distribution function $f(E,t)$ approaches thermal
equilibrium through
\begin{equation}
    f(E,t) = \frac{1}{\expp{E/T} -1} + c(E) \expp{-
    \Gamma(E) t},
\end{equation}
where $c(E)$ depends on the initial conditions. Hence the
imaginary part of the self-energy corresponds to an observable
quantity.

To order $\kappa^2$ a real particle can neither emit nor absorb a
real graviton, because these processes are kinematically
forbidden. Hence, at this order, the imaginary part of the
self-energy must vanish on shell. However in the on-shell limit
$p^0 \to E_\vect p=\sqrt{|\vect p|^2+m^2}$, we obtain, from
Eq.~\eqref{ImSigmaT},
\begin{equation} \label{ImDOnShell}
\begin{split}
    \Im \Sigma_\mathrm R(p) \xrightarrow[p^0 \to E_\vect p]{} -&\ \frac{\kappa^2 m^4T}{32\pi |\vect p|
    }
    \ln \left(\frac{{\sqrt{m^2 + |\vect p|^2}+|\vect p| }}
      {{\sqrt{m^2 + |\vect p|^2}-|\vect p|}}\right),
\end{split}
\end{equation}
which is nonzero if $T\neq 0$.

This non zero result for the on shell self-energy is an artifact
of not having introduced an infrared regularization; see
Refs.~\cite{Weldon99,Rebhan92}. To illustrate this point one can
take the on-shell limit directly in Eq.~\eqref{IntImSigma}: in
this limit $Q_1,Q_2 \to 0$, so that the phase space of the
integral vanishes while the thermal function $F(p^0,0)$ diverges.
Hence the non vanishing result we obtained is a direct consequence
of the infrared divergence of the Bose-Einstein distribution. When
regulating the infrared behavior ---for instance by giving a tiny
mass to the graviton--- no imaginary part is found  at $p^2=-m^2$.
Additionally, had we worked in an arbitrary gauge we could have
verified that the result of Eq.~\eqref{ImDOnShell} is not even
gauge invariant \cite{Weldon99,Rebhan92} \footnote{We are grateful
to A. Weldon for drawing our attention to this point.}.

\section{Discussion of the results}

At zero temperature, the position of the pole of the propagator
gives the energy of the state and hence defines the dispersion
relation, according to the K\"allen-Lehmann specral representation
\cite{WeinbergQFT,Hatfield,Peskin}. The position of the pole is
found to be
\begin{equation}
\begin{split}
    (p^0)^2  &= m^2 +|\vect p|^2 + \Sigma\TO(-m^2) = m^2 + |\vect p|^2,
\end{split}
\end{equation}
where the second equality is a consequence of the renormalization
condition \eqref{RenorCond}. The dispersion relation is clearly
Lorentz invariant, as expected.

Similarly, at finite temperature it can be shown
\cite{DonoghueEtAl85} that the ``effective'' dispersion relation
of the particle, given by the location of the poles of the
retarded propagator, determines the inertial properties of the
particle. The location of the poles is given by
\begin{equation} \label{DispRel}
     (p^0)^2 - |\vect p|^2 = m^2 + \Re \SigmaR(p^0,\vect p).
\end{equation}
The thermal mass is obtained by setting $\vect p=\vect 0$:
\begin{equation}
    \mth^2 = m^2 + \Re \SigmaR(\mth,\vect0).
\end{equation}
In a Lorentz invariant situation one would simply have $(p^0)^2 =
\mth^2+ |\vect p|^2$, but in general there can be additional
dependence on the three-momentum $\vect{p}$ on the right-hand
side,
\begin{equation} \label{RelDispTh}
    (p^0)^2 = \mth^2+|\vect p|^2 + \mathcal F(\kappa,T,\mth,\vect p).
\end{equation}
The Lorentz-breaking additional term in the dispersion relation
leads to modifications of the group velocity of the particles
$\vect v = \mathrm dp^0/\mathrm d\vect p$:
\begin{equation}\label{GroupVelocity}
    \vect v = \frac{\vect p}{p^0} + \frac{1}{2p^0}\derp{\mathcal F}{\vect
    p}.
\end{equation}

Let us find the explicit form of the thermal mass and
Lorentz-breaking terms both in the low and high temperature
regimes. Equation \eqref{DispRel} can be solved perturbatively:
\begin{equation} \label{DispRelGen}
     (p^0)^2 = m^2 + |\vect p|^2 +  \Re \Sigma_\mathrm R(E_\vect p,\vect
    p) + O(\kappa^4),
\end{equation}
where we recall that $E_\vect p = \sqrt{ m^2 + |\vect p|^2}$. At
low temperatures the modified dispersion relation, according to
Eqs.~\eqref{ReSigmaT} and \eqref{DispRelGen}, is approximately
given by
\begin{equation}
\begin{split} \label{RelDispT}
    (p^0)^2 &\approx \mth^2 +|\vect p|^2
        -   \kappa^2 \sqrt{\dfrac{ m^5 T^3
        }{{2\pi^3}}} \expp{-m/T}
        \left( \dfrac{|\vect p|^2}{3m^2+4|\vect p|^2} \right),
\end{split}
\end{equation}
where the leading contribution to the thermal mass is
\begin{equation} \label{PhysMassT}
    \mth^2 \approx m^2 - \frac{1}{6}  \kappa^2 m^2 T^2.
\end{equation}
In the high three-momentum limit the term which modifies the
dispersion relation becomes a constant and can be reabsorbed in
the thermal mass. The group velocity is modified according to
\Eqref{GroupVelocity}:
\begin{equation}
    \vect v \approx \frac{\vect p}{p^0} \left( 1 - \frac{  {\kappa }^2 m^{9/2} {T  }^{3/2} \expp{-m/T} }
    {{\sqrt{2 \pi^3}}\,{\left( 3m^2 + 4|\vect p|^2  \right)}^2}
       \right).
\end{equation}
Notice that at low temperatures the Lorentz breaking term carries
a Boltzmann factor $\expp{-m/T}$. This is due to the mentioned
fact that the nontrivial momentum dependence comes from the
thermal scalar particles whose abundance is exponentially
suppressed at low temperatures. Analogously to what happens in
electrodynamics \cite{DonoghueEtAl85}, the effect of the graviton
bath only shows up in the thermal mass.

At high temperature, $T\gg m$, the modified dispersion relation is
found to be
\begin{equation} \label{RelDispHighT}
\begin{split}
     (p^0)^2 &\approx \mth^2 + |\vect p|^2  \\ &\quad+  \frac{\kappa^2 m^2 T^2}{48}
         \left[ \frac{ E_\vect p}{|\vect p|} \ln \left(\frac{ 2E_\vect p+ |\vect p|}{ 2E_\vect p- |\vect p|}\right) - 1 \right],
\end{split}
\end{equation}
and the thermal mass is
\begin{equation}
    \mth^2 \approx m^2 - \frac{11}{48} \kappa^2 m^2 T^2.
\end{equation}
At high three-momentum the modification of the dispersion relation
can be also reabsorbed in the thermal mass. The group velocity is
given by
\begin{equation} \label{SpeedHighT}
\begin{split}
    \vect v \approx \frac{\vect p}{p^0} \bigg[ 1 &+ \frac{1}{96} \frac{
    \kappa^2 T^2 m^4}{(4m^2+3|\vect p|^2)|\vect p|^2} \\ &-
    \frac{1}{96} \frac{\kappa^2 T^2  m^4}{|\vect p|^3 E_\vect p} \ln \left( \frac{ 2
    E_\vect p + |\vect p|}{2 E_\vect p - |\vect p|} \right)
    \bigg].
\end{split}
\end{equation}
At high temperature the terms which break the Lorentz symmetry are
no longer exponentially suppressed. Notice also that Eq.
(\ref{RelDispHighT}) shows that there is no modification to the
dispersion relation for massless scalars.

The Lorentz breaking term is negative in the low-temperature case,
which implies that the speed of propagation is lowered with
respect to the standard relativistic case. In contrast, the
Lorentz breaking corrections have a positive sign at high
temperatures, so that the speed of propagation is increased.
However, it is always lower than the speed of light, as can be
seen by expanding \Eqref{SpeedHighT} in the ultrarelativistic
limit:
\begin{equation}
    \vect v \approx \frac{\vect p}{|\vect p|} \left(1 - \frac{m^2\left[ 48 + {\kappa }^2 T^2 \left( \ln 3 -1 \right)
       \right] }{96 {|\vect p|}^2} + \cdots \right)  .
\end{equation}

We should emphasize that the above breaking of the Lorentz
symmetry is only effective: if one applies a Lorentz
transformation simultaneously to the particle and the bath, all
results are Lorentz invariant. This can be seen can by introducing
the unit vector $l^\mu$ which gives the four-velocity of the
thermal bath. Then the energy and the three-momentum of the
particle with respect to the bath are
\begin{equation}
    p^0 = -l_\mu p^\mu, \qquad |\vect p| = \sqrt{(l_\mu
    p^\mu)^2+p_\mu p^\mu}\, .
\end{equation}
In this article for obvious reasons of simplicity we worked in the
bath rest frame where $l^\mu=(1,0,0,0)$.

For low values of the momentum the effective dispersion relation
can be expanded in powers of $\vect p$. For instance in the
low-temperature case, for momenta satisfying $|\vect p|<m$, the
effective dispersion relation \eqref{RelDispT} can be expanded as
\begin{equation}
\begin{split} \label{RelDispTExp}
    (p^0)^2 \approx \mth^2 +|\vect p|^2
        &-   \dfrac{\kappa^2 m^{5/2} T^{3/2}\expp{-m/T}
        }{\sqrt{2\pi^3}}  \frac{|{\vect p}|^2}{3m^2} \\ &+ \dfrac{\kappa^2 m^{5/2} T^{3/2}\expp{-m/T}
        }{\sqrt{2\pi^3}}
        \frac{2|{\vect p}|^4}{27m^4} + \cdots .
\end{split}
\end{equation}
It is worth mentioning the difference between this expansion and
that of Eq.~\eqref{Eq1}. In a fundamental approach to the Lorentz
symmetry breaking one expects each additional power of momentum to
be suppressed by increasing powers of the Planck mass, whereas in
the effective breaking approach we have pursued there are several
energy scales. Hence there is more freedom in the possible values
of the suppression factor. For instance, in the low-temperature
regime ---\ie, in \Eqref{RelDispTExp}--- the $|\vect p|^4$ term is
suppressed by $(T/m)^{3/2}e^{-m/T} \ll 1$ in Planck units. The
fact that in the latter approach there are more energy scales also
explains why modifications to the $|\vect p|^2$ term are present
in Eq.~\eqref{RelDispTExp} while they are not included in
Eq.~\eqref{Eq1}. Notice also that the corrections that we found
only contain even powers of the momentum, and thus are in
agreement with the results of Ref.~\cite{Lehnert03}.

In the present universe the effects we discussed are completely
negligible when applied to electrons or protons, both because they
are proportional to the Planck length square and because they are
exponentially suppressed. To obtain relevant effects, one should
consider Planckian temperatures. However, in this case
perturbation theory around a flat-spacetime approximation probably
fails. Therefore the results obtained here should be considered an
indication that effective violations of Lorentz invariance indeed
occur, and that gravitational interactions cannot be neglected
when exploring highly energetic regions of dispersion relations.

\begin{figure}
    \hfill
    \includegraphics{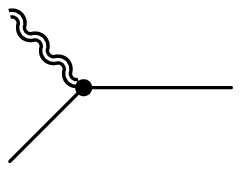}
    \hfill
    \includegraphics{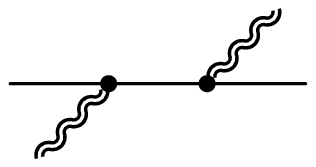}
    \hfill\mbox{}
    \caption{Diagrams such as the one depicted on the left would
    contribute to order $\kappa^2$ to the thermalization rate at finite temperature, and hence to the
    imaginary part of the self-energy. However all those one-vertex
    diagrams are identically zero because of kinematical reasons.
    The first non-vanishing contribution would be due to
    Compton-like diagrams such as the one depicted on the right,
    and hence it would be a $\kappa^4$ contribution.}
    \label{fig:cut}

\end{figure}

Whereas the real part of the self-energy gives the change in
energy of the particle and hence the dispersion relation, the
imaginary part accounts for the dissipative effects and gives the
thermalization rate. As we have already commented, at the order
$\kappa^2$ the imaginary part of the self-energy is zero. We do
not see the thermalization effect since all processes which could
contribute to it, such as the one shown in Fig.~\ref{fig:cut} (on
the left), are kinematically forbidden. In order to account for
thermalization effects we should compute the imaginary part of the
self-energy to order $\kappa^4$. At this order, the particle can
exchange momentum with the thermal bath through processes such as
the one to the right of Fig.~\ref{fig:cut}, which would correspond
to the Compton scattering in electrodynamics. Notice here a
significant difference with respect to the vacuum case: while at
zero temperature the self-energy is purely real to all orders of
perturbation theory (because the particle is stable), at finite
temperature it will acquire an imaginary contribution to order
$\kappa^4$.

Let us end with a short summary of the main points of the paper.
We have illustrated with a particular example how local Lorentz
symmetry is effectively violated because of the interactions with
a nontrivial ensemble of metric fluctuations, even if Lorentz
symmetry holds at a fundamental level. We have also shown how this
effective violation can be addressed at low energies within
standard physics. The main quantitative result of the paper are
Eqs.~\eqref{RelDispT} and \eqref{RelDispHighT}, which explicitly
show the modifications of the dispersion relation. As in the
electromagnetic case, this effect is exponentially suppressed at
low temperatures. Moreover, the modifications of the dispersion
relation are suppressed when the three-momentum of the massive
particle (defined in the heat bath rest frame) is much larger than
the temperature and the mass. This last result is somewhat
unexpected since the gravitational coupling grows with the energy,
unlike the electromagnetic case. Therefore, in spite of the
derivative coupling, no violation of Lorenz invariance is found in
the high-momentum limit, at least at one loop. Finally we have
also shown that gravitational interactions do not generate a
four-derivative term in the scalar field action.

\begin{acknowledgments}
    We would like to thank E. Calzetta, J. F. Donoghue, X. Garcia i Tormo,
    B.-L. Hu, A. Roura and A. Weldon for their useful suggestions
    and for interesting discussions during various stages of this project.
    We are grateful to E.~Gunzig for the organization of
    the Peyresq Physics meetings which stimulated
    discussions on this topic. D.~A. acknowledges
    support of a FI grant from the Generalitat de Catalunya. This
    work is also partially supported by MICYT Research Project No.~FPA-2001-3598 and European Project No.~HPRN-CT-2000-00131.
\end{acknowledgments}

\appendix

\section{Feynman rules} \label{app:FeyRules}

At zero temperature the free propagator for the scalar field is
\begin{equation}
    G_\mathrm F^{(0,{\scriptscriptstyle T=0})}(p)=\frac{-i}{p^2+m^2-i\epsilon},
\end{equation}
and the free propagator for the gravitons in the harmonic gauge is
\cite{Donoghue94a,Donoghue94b,Veltman76}
\begin{equation}
    (\Delta_\mathrm F)\TO_{\mu\nu\alpha\beta}(p) = \frac{-i \mathcal P_{\mu\nu\alpha\beta}}{p^2-i\epsilon},
\end{equation}
where
\begin{equation}
    \mathcal P_{\mu\nu\alpha\beta} = \fud \left(
    \eta_{\mu\alpha}\eta_{\nu\beta} + \eta_{\mu\beta}\eta_{\nu\alpha}
    - \frac{2}{d-2} \eta_{\mu\nu}\eta_{\alpha\beta}
    \right)
\end{equation}
in $d$ spacetime dimensions. In four dimensions, $\mathcal
P_{\mu\nu\alpha\beta} =  \left(
    \eta_{\mu\alpha}\eta_{\nu\beta} + \eta_{\mu\beta}\eta_{\nu\alpha}
    - \eta_{\mu\nu}\eta_{\alpha\beta}
    \right)/2$.

\begin{figure}
    \centering
    \hfill
    \includegraphics{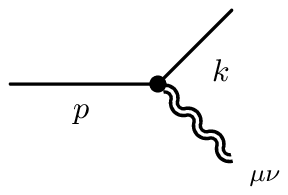}
    \hfill
    \includegraphics{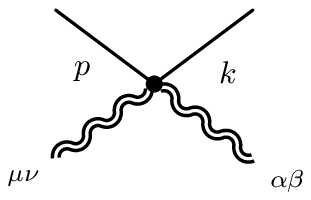}
    \hfill\mbox{}
    \caption{On the left, the two-scalar--one-graviton vertex, denoted
    by $\tau_{\mu\nu}(p)$. On the right, the two-scalar--two-graviton vertex,
    denoted by $V_{\mu\nu\alpha\beta}(p)$.}
    \label{fig:Vert}
\end{figure}

The two-scalar--one-graviton vertex, shown in Fig.~\ref{fig:Vert},
is given by \cite{Donoghue94b}
\begin{equation}
    \tau_{\mu\nu}(p,k) = \frac{i \kappa}{2} [ p_\mu k_\nu + p_\nu k_\mu
    -
    (p\cdot k) \eta_{\mu\nu} - m^2 \eta_{\mu\nu} ],
\end{equation}
and the two-scalar--two-graviton vertex, also shown in
Fig.~\ref{fig:Vert}, is given by \cite{Donoghue94b}
\begin{equation}
\begin{split}
    V_{\mu\nu\alpha\beta}(p,k) = - &\ i\kappa^2  \bigg[
    I_{\mu\nu\rho\lambda}
    I^\lambda_{\phantom\lambda\sigma\alpha\beta} \left(p^\rho k^\sigma
    + p^\sigma k^\rho\right) \\
    &+\frac12 \left(\eta_{\mu\nu} I_{\alpha\beta\rho\sigma} +
    \eta_{\alpha\beta} I_{\mu\nu\rho\sigma} \right) p^\rho
    k^\sigma \\
    &-\frac12 \left(I_{\mu\nu\alpha\beta} - \fud
    \eta_{\mu\nu}\eta_{\alpha\beta}\right)(p\cdot k + m^2) \bigg],
\end{split}
\end{equation}
where
\begin{equation*}
    I_{\mu\nu\alpha\beta} = \fud \left(
    \eta_{\mu\alpha}\eta_{\nu\beta} +
    \eta_{\mu\beta}\eta_{\nu\alpha}\right).
\end{equation*}

At finite temperature the free propagators for the scalar field
are given by Eq.~\eqref{FreeGT} and the free propagators for the
gravitons are $(\Delta_{ab})_{\mu\nu\alpha\beta}(p)=\mathcal
P_{\mu\nu\alpha\beta} G^{(0)}_{ab}(p)|_{m=0}$. Vertices type 1 are
described by $\tau_{\mu\nu}$ and $V_{\mu\nu\alpha\beta}$ and
vertices type 2 are described by $-\tau_{\mu\nu}$ and
$-V_{\mu\nu\alpha\beta}$.

\section{Integrals in dimensional regularization}\label{app:Int}

In Eq.~\eqref{SigmaDecInt} we need the following integrals:
\begin{gather}
I(p) = \mu^\varepsilon\int \frac{\vd[d] k}{(2\pi)^d}
     \frac{1}{[k^2+m^2-i\epsilon][(p-k)^2-i\epsilon]}, \\
I_\mu(p) = \mu^\varepsilon\int \frac{\vd[d] k}{(2\pi)^d}
    \frac{k_\mu}{[k^2+m^2-i\epsilon][(p-k)^2-i\epsilon]},\\
I_{\mu\nu}(p) = \mu^\varepsilon \int \frac{\vd[d] k}{(2\pi)^d}
\frac{k_\mu
    k_\nu}{[k^2+m^2-i\epsilon][(p-k)^2-i\epsilon]}.
\end{gather}
The result in arbitrary $d$ dimensions after series expansion in
$\varepsilon = 4-d$ is
\begin{widetext}
\begin{align}
    I &= \frac{i}{(4\pi)^2} \left[ \frac{1}{\hat\varepsilon} +  2
    - \frac{m^2}{p^2} \ln \left(1+\frac{p^2}{m^2} -i\epsilon \right) - \ln \left(\frac{p^2+m^2}{\mu^2}-i\epsilon\right)\right] +
    O(\varepsilon),\\
  I_\mu(p)&= \frac{i p_\mu}{2(4\pi)^2} \bigg[ \frac{1}{\hat\varepsilon} +  2 -
    \frac{m^2}{p^2}
    + \frac{m^4}{p^4} \ln \left(1+\frac{p^2}{m^2} -i\epsilon\right)
    -  \ln \left(\frac{p^2+m^2}{\mu^2}-i\epsilon\right)\bigg] +
    O(\varepsilon),\\
\begin{split}
    I_{\mu\nu}(p) &= - \frac{\eta_{\mu\nu}}{2(4\pi)^2} \bigg[ \frac{m^2}{2\hat\varepsilon} + \frac{p^2}{6\hat\varepsilon}+ \frac{m^4}{6p^2} + \frac{7m^2}{6}
    +
    \frac{4p^2}{9}\\
      &\qquad - \left( \frac{m^6}{6p^4} + \frac{m^4}{2p^2} \right) \ln \left( 1 + \frac{p^2}{m^2}-i\epsilon \right)  -\left(\frac{m^2}{2}+ \frac{p^2}{6}\right)\ln \left( \frac{p^2+m^2}{\mu^2} - i\epsilon \right) \bigg]
    \\
    &\quad+ \frac{p_\mu p_\nu}{(4\pi)^2}
    \bigg[ \frac{1}{3\hat\varepsilon}  - \frac{m^6}{3p^6}
    \ln\left(1+\frac{p^2}{m^2}-i\epsilon
    \right)-\frac13  \ln\left(\frac{p^2+m^2}{\mu^2}-i\epsilon\right)
     + \frac{m^4}{3p^4} - \frac{m^2}{6p^2} + \frac{13}{18} \bigg] +
     O(\varepsilon),
\end{split}
\end{align}
where
\[
\frac{1}{\hat\varepsilon} = \frac{2}{\varepsilon} - \gamma +
    \ln 4\pi.
\]
\end{widetext}

\section{Real-time thermal field theory}\label{app:TFT}

For the benefit of those readers who might be unfamiliar with the
subject we give in this appendix a brief account of those aspects
of real-time thermal field theory which are needed in the paper.
For a complete introduction to the thermal field theory we address
the reader to Refs.~\cite{LandsmanWeert87,DasThermal,LeBellac};
see also Ref.~\cite{CamposHu98} for an approach similar
\nopagebreak to ours.

\subsection{Correlation functions}

\begin{figure}
    \centering
    \includegraphics{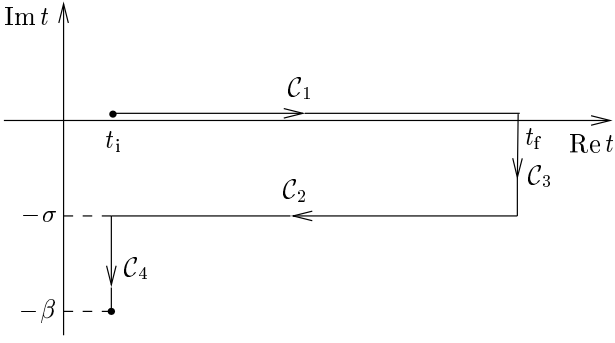}
    \caption{Integration contour in the complex-time plane used in the
    real-time approach to thermal field theory. The choice
    $\sigma=0^+$ makes the formalism analogous to the CTP approach to
    field theory.}
    \label{fig:Path}
\end{figure}

In thermal field theory the initial state for the fields is
assumed to be a thermal state at temperature $T=1/\beta$, which is
characterized by a density matrix
\begin{equation}
     \hat\rho = \frac{\expp{- \beta\hat
    H}}{\Tr(\expp{- \beta\hat {H}})},
\end{equation}
where $ \hat{H}$ is the Hamiltonian operator of the system (for
the purposes of this appendix we shall consider a scalar field
$\phi$). The generating functional $Z_{\mathcal C}[j]$ is defined
as
\begin{equation}
    Z_{\mathcal C}[j] = \Tr \left(\hat \rho T_{\mathcal C} \expp{i \int_{\mathcal C} \vd t \int
    \ud[3]{\vect
    x} \hat \phi(x) j(x)}  \right),
\end{equation}
where $\hat \phi(x)$ is the field operator in the Heisenberg
picture, ${\mathcal C}$ is a certain path in the complex $t$
plane, $T_{\mathcal C}$ means ordering along this path and $j(x)$
is a classical external source. By functional differentiation of
the generating functional with respect to $\phi$, path-ordered
correlation functions can be obtained. The generating functional
may be computed in a path-integral representation as
\begin{equation}
    Z_{\mathcal C}[j] = \int \uD \phi \expp{ i  \int_{\mathcal C} \vd t \int
    \ud[3]{\vect x} \left\{ {\mathcal L}[\phi(x)] + j(x) \phi(x)\right\} },
\end{equation}
where ${\mathcal L}$ is the Lagrangian density and the contour
${\mathcal C}$ is now restricted to begin at some initial real
time $t_\mathrm i$ and to end at $t_\mathrm i - i\beta$. The
boundary conditions for the path integral are $\phi(t_\mathrm
i,\vect x) = \phi(t_\mathrm i-i\beta,\vect x)$. Different
elections for ${\mathcal C}$ lead to different approaches to
thermal field theory: a straight line from $t_\mathrm i$ to
$t_\mathrm i - i\beta$ leads to the imaginary-time formalism, and
the contour shown in Fig.~\ref{fig:Path} leads to the real-time
formalism. By choosing $\sigma= 0^+$ our formalism will coincide
with the closed time path (CTP) approach to non equilibrium field
theory \cite{Schwinger61,Keldysh65,ChouEtAl85}. In fact many of
the properties listed below are not limited to the thermal case
but are valid in a more general non equilibrium situation. We
shall try to specify in each case which properties are general and
which ones are particular to the thermal case.

If we are interested in real-time correlation functions, the path
along ${\mathcal C}_3$ and ${\mathcal C}_4$ can be usually
neglected once we properly take into account the boundary
conditions of the path integral. If we define $\phi_{1,2}(t,\vect
x)=\phi(t,\vect x)$ and $j_{1,2}(t,\vect x)=j(t,\vect x)$ for $t
\in {\mathcal C}_{1,2}$, and take $t_\mathrm i \to -\infty$ and
$t_\mathrm f\to \infty$ the generating functional can be
reexpressed as $Z_{\mathcal C}[j] = Z_{34} Z[j_1,j_2]$, where
$Z_{34}$ represents the constant contribution from the segments
${\mathcal C}_3$ and ${\mathcal C}_4$ and $Z[j_1,j_2]$ is
\begin{equation}
\begin{split}
    Z[j_1,j_2]= \int \uD \phi_1 \uD \phi_2 & \expp{i\int \ud[4]x
    \left\{ {\mathcal L}[\phi_1(x)] + j_1(x) \phi_1(x)\right\}}
    \\ \times & \expp{ -i\int \ud[4]x \left\{
      {\mathcal L}[\phi_2(x)] + j_2(x)
    \phi_2(x) \right\}}.
\end{split}
\end{equation}
Correlation functions are defined by second functional
differentiation with respect to the external source $j(x)$:
\begin{equation}
    G_{ab}(x,x') =
    \left.\frac{1}{Z[0,0]}\derff{Z[j_1,j_2]}{j^a(x)}{j^b(x)}\right|_{j_1=j_2=0},
\end{equation}
where $a,b = 1,2$. In operator language correlation functions can
be written as
\begin{subequations} \label{CorrFunct}
\begin{align}
    G_{11}(x,x') &= \av{ T \hat\phi(x) \hat\phi(x') }, \\
    G_{12}(x,x') &= \av{ \hat\phi(x) \hat\phi(x') }, \\
    G_{21}(x,x') &= \av{ \hat\phi(x') \hat\phi(x) }, \\
    G_{22}(x,x') &= \av{ \widetilde T \hat\phi(x) \hat\phi(x') },
\end{align}
\end{subequations}
where $(\widetilde T)$ $T$ is the (anti-) time-ordering operator,
and where the average means $\av{\cdots}= \Tr{(\hat\rho \cdots)}$.
Lowercase roman indices are raised and lowered with the ``metric''
$c_{ab}=\mathrm{diag}(1,-1)$. From the above expressions it can be
readily seen that not all correlation functions are independent.
The following relations in Fourier space are a consequence of
Eqs.~\eqref{CorrFunct}:
\begin{subequations} \label{RelG}
\begin{gather}
    G_{11}(p) = G_{22}^*(p), \qquad G_{12}(p)=G_{21}(-p),\\
    G_{11}(p) + G_{22}(p) = G_{12}(p) + G_{21}(p),
\end{gather}
\end{subequations}
where correlation functions in Fourier space are
\begin{equation*}
    G_{ab}(p) = \int \ud[4]{x} \expp{-i p \cdot \Delta}
    G_{ab}(X+\Delta/2,X-\Delta/2),
\end{equation*}
where we have introduced the new variables $\Delta=x-x'$ and
$X=(x+x')/2$ \footnote{In order to avoid cumbersome notation the
same symbol is used for correlators in configuration space and in
Fourier space. Notice that Fourier-transformed propagators depend
on $X$ if the initial state is non homogeneous. Since our primary
concern here is thermal states, which are homogeneous, we do not
indicate explicitly this $X$ dependence.}. Additionally, in the
case of an initial thermal state the following relation, a
consequence of the Kubo-Martin-Schwinger (KMS) formula (equivalent
to the fluctuation-dissipation theorem, in another context), is
also verified:
\begin{equation}
    G_{11}(p) + G_{22}(p) = \expp{-\beta p^0} G_{12}(p) + \expp{\beta p^0}
    G_{21}(p).
\end{equation}
Thus, in thermal field theory the knowledge of just one
correlation function determines all of them.

An important correlation function is the retarded propagator,
which is defined as
\begin{equation}
    G_\mathrm R(x,x') = \theta(x^0-x'^0) \av{[\hat
    \phi(x),\hat\phi(x')]},
\end{equation}
and is related to the other correlation functions through
\begin{equation}
    G_\mathrm R(p) = G_{11}(p) - G_{12}(p).
\end{equation}
The retarded propagator has the remarkable property that has well
defined analyticity properties at finite temperature, as opposite
to most other propagators: it is analytic in the upper half of the
complex $p^0$ plane. Furthermore, the retarded propagator is the
one that one naturally obtains from an analytic continuation of
the Euclidian propagator in the imaginary-time formalism
\cite{FetterWalecka}.

\subsection{Perturbation theory}

Perturbation theory can be organized in a similar way as in the
zero-temperature case, but taking into account that at finite
temperature there are two kind of vertices ($1$ and $2$) and four
kind of propagators ($11$, $12$, $21$ and $22$) which link the two
vertices. Vertices type 2 carry an additional minus sign with
respect to vertices type 1. When computing Feynman diagrams one
has to sum over all possible internal vertices. For a real scalar
field of mass $m$, free propagators are given by
\begin{equation} \label{FreeGT}
\begin{split}
         G_{ab}^{(0)}(p)  &=
        \begin{pmatrix}
            \dfrac{-i}{p^2 + m^2 - i \epsilon} & {2\pi \delta(p^2+m^2) \theta(-p^0)}\\
            { 2\pi \delta(p^2+m^2) \theta(p^0)}
             & \dfrac{i}{p^2 + m^2 + i \epsilon}
        \end{pmatrix} \\ &\quad + 2\pi \delta(p^2+m^2) {n(|p^0|)}
        \begin{pmatrix}
            1 & 1 \\
            1 & 1
        \end{pmatrix},
\end{split}
\end{equation}
where $n(E)$ is the Bose-Einstein distribution function:
\begin{equation}
    n(E) = \frac{1}{1-\expp{\beta E}}.
\end{equation}
Thermal contributions to Feynman diagrams are always finite in the
ultraviolet regime because the Bose-Einstein function acts as a
soft cutoff for momenta larger than the temperature $T$. The
counterterms which renormalize the theory at zero temperature also
renormalize the theory at finite temperature. Note also that the
thermal part of the propagator, which breaks the Lorentz symmetry
through an explicit dependence on $p^0$, is always on shell.

Instead of working with the four propagators $G_{ab}(p)$ in the
thermal case one can reorganize perturbation theory in a way such
that just retarded and advanced propagators are involved
\cite{EijckKobesWeert94}. In this case one has to consider how
Feynman rules are transformed when working with the
retarded/advanced basis.

\subsection{Self-energy}

In a thermal or, more generally, in a non equilibrium situation,
the self-energy has a matricial structure and is implicitly
defined through the equation
\begin{equation} \label{SDM}
    G_{ab}(p) = G_{ab}^{(0)}(p)+
    G_{ac}^{(0)}(p) [-i\Sigma^{cd}(p)]  G_{db}(p),
\end{equation}
where $G^{(0)}_{ab}(p)$ are the free propagators of the theory.
The $ab$ component of the self-energy can be computed, similarly
to the vacuum case, as the sum of all one-particle irreducible
diagrams with  amputated external legs that begin and end with
type $a$ and type $b$ vertices, respectively.

In general the self-energy components verify the non-perturbative
relations
\begin{subequations} \label{RelSigma}
\begin{gather}
    \Sigma^{11}(p) =- (\Sigma^{22})^*(p), \qquad \Sigma^{12}(p)=\Sigma^{21}(-p)\\
    \Sigma^{11}(p) + \Sigma^{22}(p) =-  \Sigma^{12}(p) -
    \Sigma^{21}(p),
\end{gather}
\end{subequations}
which can be obtained from Eqs.~\eqref{RelG} and \eqref{SDM}. The
following equation is just verified if the initial state is
thermal:
\begin{equation}
    \Sigma^{11}(p) + \Sigma^{22}(p) = -\expp{-\beta p^0} \Sigma^{12}(p) - \expp{\beta p^0}
    \Sigma^{21}(p).
\end{equation}
Thus, all the components of the self-energy can be determined from
knowledge of just one of them. Combining relations
\eqref{RelSigma} we obtain
\begin{equation} \label{cut}
    \Im \Sigma^{11}(p) = \frac{i}{2} [ \Sigma^{12}(p) +
    \Sigma^{21}(p)
    ].
\end{equation}
This last equation can be directly obtained from the cutting rules
at finite temperature.

A particularly useful combination is the  retarded self-energy,
defined as $\Sigma_\mathrm R(p) = \Sigma^{11}(p) +
\Sigma^{12}(p)$. It  is related to the retarded propagator through
\begin{equation}\label{SigmaAR}
    {G}_\mathrm{R}(p) = \frac{-i}{p^2 + m^2+
    \Sigma_\mathrm{R}(p)}.
\end{equation}
The above relation, which justifies the name of retarded
self-energy for $\Sigma_\mathrm{R}(p)$, can be demonstrated by
expanding the matrix equation \eqref{SDM} and using the relations
\eqref{RelG} and \eqref{RelSigma}. Similar relations hold for the
advanced propagator $G_\mathrm A(p) = G^*_\mathrm R(p)$ and the
advanced self-energy $\Sigma_\mathrm A(p) = -\Sigma^*_\mathrm
R(p)$. Notice that a diagonal relation such as Eq.~\eqref{SigmaAR}
can be found only for the retarded (or advanced) propagator.

According to \Eqref{cut}, the imaginary part of the retarded
self-energy can be also expressed as
\begin{equation} \label{cutR}
    \Im \Sigma_\mathrm R(p) = \frac{i}{2} [ \Sigma^{21}(p) -
    \Sigma^{12}(p)
    ].
\end{equation}
For a thermal state $\Sigma_\mathrm R(p)$ is related to
$\Sigma^{11}(p)$ through:
\begin{equation}
    \Sigma^\mathrm R(p)  = \Re \Sigma^{11}(p) + \tanh
    \left( \frac{p^0}{2T} \right) \Im \Sigma^{11}(p). \label{SigmaR11}
\end{equation}

\section{Computation of $A(p)$, $B(p)$, $C(p^2)$ and
$D(p)$}\label{app:ABCD}

In this appendix we shall compute the integrals $A(p)$, $B(p)$,
$C(p^2)$ and $D(p)$ which appear in the calculation of the
self-energy at finite temperature; see Eqs.~\eqref{A}--\eqref{C}
and \eqref{D}.

Let us start by computing the integral $A(p)$, defined in
\Eqref{A}. The Dirac delta can be expanded as
\begin{equation*}
    \delta\boldsymbol((p-k)^2\boldsymbol) = \delta(q^2)
    = \frac{1}{2|\vect q|} \left[ \delta(-q^0+ |\vect q|) +
    \delta(q^0+|\vect{q}|) \right],
\end{equation*}
where we have introduced the new variable $q=p-k$ and where
$p=(p^0,\vect p)$ and $q=(q^0,\vect q)$. Introducing now spherical
coordinates $(\phi,\theta)$ in the three spatial dimensions, with
$\theta$ being the angle between $\vect p$ and $\vect q$, and
integrating with respect to $q^0$ with the aid of the delta
function we get
\begin{widetext}
\begin{equation*}
    A(p) = \int_0^\infty \frac{n(Q) Q\mathrm d Q}{2(2\pi)^2} \int_{-1}^1 \ud x  \left[ \PV\frac{
     g_2(p^2,0,-p^0 Q +  P Q x)}{-(p^0)^2+P^2+2p^0 Q -  2P Q x+m^2} \right. + \left. \PV\frac{
     g_2(p^2,0,p^0 Q +  P Q x)}{-(p^0)^2+P^2-2p^0 Q - 2 P Q x+m^2} \right],
\end{equation*}
where $Q=|\vect q|$, $P=|\vect p|$, $x=\cos \theta$,
\begin{equation}
\begin{split}
    g_2(p^2,q^2,p\cdot q)&=g_1\boldsymbol(p^2,(p-q)^2,p\cdot(p-q)\boldsymbol)
    = -2m^4 - 2 m^2 p^2 + p^4 + 2 m^2 (p \cdot q) - 2 p^2 (p \cdot q)+
      p^2 q^2,
\end{split}
\end{equation}
and we have performed the trivial angular integration over $\phi$.
We now integrate with respect to $x$ to get
\begin{equation} \label{PreB}
\begin{split}
    A(p) = \int_0^\infty \frac{n(Q) \mathrm d Q}{16\pi^2P} &\
    \Bigg[
    8PQ[-m^2-(p^0)^2+P^2] + m^2\{m^2+2[(p^0)^2-P^2]\} \\  &\quad\times\ln
    \left(
    \frac{\left[m^2 - (p^0-P+2Q)(p^0+P)\right]\left[m^2 -
    (p^0-P)(p^0+P-2Q)\right]}{\left[m^2 - (p^0-P-2Q)(p^0+P)\right]\left[m^2 -
    (p^0-P)(p^0+P+2Q)\right]}\right) \Bigg].
\end{split}
\end{equation}
\end{widetext}
The result of this integral cannot be given in closed analytic
form, in general. However in this paper we are mainly interested
in its on-shell value $p^0 = E_\vect p = \sqrt{m^2+P^2}$, and in
this limit the integral can be computed exactly at any temperature
since the logarithmic term in \Eqref{PreB} vanishes. In this case
the value of the integral is given by
\begin{equation} \label{AOnShell}
    A(E_\vect p,\vect p) = - \frac{ m^2}{\pi^2} \int_0^\infty \ud Q n(Q)
    Q = - \frac{1}{6}m^2T^2,
\end{equation}
where we used that
\begin{equation*}
    \int_0^\infty \ud Q n(Q) Q = \frac{\pi^2 T^2}{6}.
\end{equation*}

We now proceed with the computation of $B(p)$, defined in
\Eqref{B}. Repeating similar steps as in the previous integral we
get
\begin{widetext}
\begin{equation*}
\begin{split}
    B(p) = \int_0^\infty \frac{n(E_\vect k) K^2\mathrm d K}{2(2\pi)^2E_\vect k} \int_{-1}^1 \ud x  \left[ \PV\frac{
     g_1(p^2,0,-p^0 E_\vect k +  P K x)}{-(p^0)^2+P^2+2p^0 E_\vect k - 2 P K x} \right. + \left. \PV\frac{
     g_1(p^2,0,p^0 E_\vect k +  P K x)}{-(p^0)^2+P^2-2p^0 E_\vect k - 2 P K x}
     \right],
\end{split}
\end{equation*}
\end{widetext}
where we recall that $E_\vect k=\sqrt{m^2+K^2}$. The integral with
respect to $x$ can be analytically performed; the result is a very
large and cumbersome expression, which we shall not reproduce
here. The resulting expression cannot be integrated again in a
closed analytic form. However, we may find particular expressions
valid at low and high temperatures. As in the case of previous
integral, we will restrict to the on shell results.

At low temperatures, only those momenta $\vect k$ whose
corresponding energies are at most of the order of the
temperature, $E_\vect k \lesssim T$, contribute significatively to
the integral because of the presence of the thermal factor
$n(E_\vect k)$, which acts as a soft cutoff. Hence, low
temperature also implies low-energy and low momentum. Therefore,
in the low-temperature approximation we may retain only the
leading term in a $K$ expansion:
\begin{equation*}
\begin{split}
    B(E_\vect p, \vect p) &=  \frac{m^2+2P^2}{\pi^2(3m^2+4P^2)} \\&\quad\times
    \int_0^\infty \ud K n(E_\vect k) \left[ K^2 +
    O(K^3) \right].
\end{split}
\end{equation*}
Taking into account that for low temperatures
\begin{equation*}
    n(E_\vect k) \approx
    \expp{-m/T}\expp{-K^2/(2mT)}
\end{equation*}
and that
\begin{equation*}
    \int_0^\infty \ud K \expp{-K^2/(2mT)} K^2 = \sqrt{\frac\pi2}
    \,
    (mT)^{3/2},
\end{equation*}
we find the following expression for $B(p)$ at low temperature:
\begin{equation}
     B(E_\vect p,\vect p) \approx \sqrt{\frac{m^5 T^3}{2\pi^3}}
     \left(\frac{m^2+2P^2}{3m^2+4P^2}\right)\expp{-m/T}.
\end{equation}
We have not made precise the exact meaning of the
``low-temperature'' approximation employed above. In principle
this approximation would require the temperature $T$ to be much
smaller than any relevant quantity with dimensions of energy that
could be formed by a combination of $m$ and $P$. However, a
detailed analysis of the expressions shows that the condition $m
\gg T$ is sufficient to guarantee the validity of the result.

Let us now proceed to the calculation of $B(p)$ in the high
temperature regime. Since $B(p)$ would be divergent if no thermal
cutoff were present, at high temperatures the leading contribution
to the integral is given by those momenta close to the temperature
$T$.Thus as a first approximation we can retain only the leading
term in a $1/K$ expansion:
\begin{equation}
\begin{split}
    B(E_\vect p, \vect p) &=
    \bigg[ \frac{m^2 \sqrt{m^2+ P^2}}{4\pi^2P} \ln \left(\frac{ 2\sqrt{m^2+ P^2}- P}{ 2\sqrt{m^2+ P^2}+ P}\right) \\
    &\quad  + \frac{3}{4\pi^2 } m^2 \bigg] \int_0^\infty \ud K
    n(E_\vect k) \left[ K + O(1/K^0) \right]
\end{split}
\end{equation}
Since the leading contribution to the integral is given in the
ultrarelativistic regime, we can approximate the energy by the
momentum in the Bose-Einstein function, $n(E_\vect k)\approx
n(K)$. With this approximation we find
\begin{equation}\label{BAppendix}
\begin{split}
    B(E_\vect p, \vect p) &=
    \frac{m^2 T^2 E_{\vect p}}{24P} \ln \left(\frac{ 2E_{\vect p}- P}{ 2E_{\vect P}+ P}\right)  + \frac{1}{8} m^2 T^2.
\end{split}
\end{equation}
Analogously to the low-temperature case, the high temperature
approximation would \emph{a priori} require the temperature $T$ to
be much higher than $m$, $P$ and any relevant energy scale formed
by combination of these two. Again, one can show that the
condition $T\gg m$ is sufficient to guarantee the validity of
\Eqref{BAppendix}.

We now move to the integral $C(p^2)$, defined in \Eqref{C}. Its
evaluation is straightforward:
\begin{equation}
    C(p^2) = g_3(p^2) \int_0^\infty \frac{K \vd K}{2\pi^2} n(K) =
    \frac{T^2}{12} (10m^2+4p^2).
\end{equation}
We only need its on shell value $C(-m^2) = T^2 m^2 / 6$.

Let us now consider the integral $D(p)$, defined in \Eqref{D}. We
start by introducing the variable $q = p -k$:
\begin{widetext}
\begin{equation*}
    \begin{split}
    D(p) = \int \frac{\mathrm d q^0 \, \mathrm d^3 \vect q}{(2\pi)^2} &\ \, F(p^0,q^0)\, g_2\boldsymbol(-(p^0)^2+|\vect p|^2,-(q^0)^2+\vect q^2,-p^0 q^0+\vect p \cdot \vect q \boldsymbol)
      \\ &\times
      \delta\boldsymbol( - (q^0)^2 + \vect q^2 \boldsymbol)\, \delta\boldsymbol(-(p^0-q^0)^2+(\vect p - \vect q)^2
    +m^2\boldsymbol),
    \end{split}
\end{equation*}
where $p=(p^0,\vect p)$ and $q=(q^0,\vect q)$. Next we expand the
first delta function and integrate over $q^0$:
\begin{equation} \label{DpNice}
    \begin{split}
    D(p) &= \int \udpi[3]{\vect q} \frac{2\pi}{2 |\vect q|} F(p^0,|\vect q|) \, g_2\boldsymbol(|\vect p|^2-(p^0)^2,0,\vect p \cdot \vect q - p^0 |\vect q|
    \boldsymbol)\,
    \delta\boldsymbol(-(p^0-|\vect q|)^2+(\vect p - \vect q)^2
    +m^2\boldsymbol)\\
    &\quad+\int \udpi[3]{\vect q} \frac{2\pi}{2 |\vect q|} F(p^0,-|\vect q|)\, g_2\boldsymbol(|\vect p|^2-(p^0)^2,0,\vect p \cdot \vect q + p^0 |\vect q| \boldsymbol) \,  \delta\boldsymbol(-(p^0+|\vect q|)^2+(\vect p - \vect q)^2
    +m^2\boldsymbol).
    \end{split}
\end{equation}
We now introduce spherical coordinates over $\vect q$ and expand
the second delta function:
\begin{equation*}
    \begin{split}
    D(p) &= \frac{1}{8\pi P} \int_{-1}^1 \ud x \int_0^\infty \ud Q F(p^0,Q) \, g_2\boldsymbol(P^2-(p^0)^2,0,Q P x - Q p^0\boldsymbol) \, \delta\boldsymbol( x - (p^2+ 2 p^0 Q + m^2
    )/(2PQ)\boldsymbol)\\
    &\quad+\frac{1}{8\pi P} \int_{-1}^1 \ud x \int_0^\infty \ud Q F(p^0,-Q) \, g_2\boldsymbol(P^2-(p^0)^2,0,Q P x + Q p^0\boldsymbol) \,  \delta\boldsymbol( x - (p^2- 2 p^0 Q + m^2
    )/(2PQ)\boldsymbol).
    \end{split}
\end{equation*}
\end{widetext}
  where $Q = |\vect q|$, $P=|\vect p|$ and $x=\cos
\theta$. We have already performed the trivial angular integration
over $\phi$. Integrating with respect to $x$ with the aid of the
delta function we get
\begin{equation}
    D(p) = \frac{g_2\boldsymbol(p^2,0,(p^2+m^2)/2\boldsymbol)}{8\pi P} \left| \int_{Q_1}^{Q_2}  \ud Q
    F(p^0,Q)\right|
\end{equation}
with
\begin{equation*}
    Q_2=\frac{(p^0)^2-P^2-m^2}{2(p^0-P)}, \qquad
    Q_1=\frac{(p^0)^2-P^2-m^2}{2(p^0+P)},
\end{equation*}
which can be finally arranged as
\begin{equation}
    D(p) = \frac{-m^2(m^2+2p^2)}{8\pi P} \left| \int_{Q_1}^{Q_2}  \ud Q
    F(p^0,Q)\right|.
\end{equation}
Recall that $p^2 = -(p^0)^2+P^2$.

\end{document}